\documentclass{aa}

\usepackage{graphicx}

\usepackage{latexsym,amsmath,amssymb}

\usepackage{hyperref}

\usepackage{txfonts}

\usepackage{natbib}

\begin{document}

\title{Dark Matter and MOOCs}

\author{Paolo Salucci\inst{1}, Claudia Antolini\inst{1}}

\institute{SISSA/ISAS, via Beirut 2-4, 34013 Trieste, Italy}

\abstract
{}
{To teach the topic of Dark Matter in Galaxies to undergraduate
and PhD students is not easy, one reason being that the 
scientific community has not converged yet to a generally 
shared knowledge. 
}
{ We argue that the teaching of this topic 
and its subsequent  scientific progress may benefit by Massive Online 
and Open Courses. }
{  The reader of this paper can express his/her opinion on 
this by means of a confidence vote    }
{  Will MOOC and Astrophysics meet?}

\keywords{ Dark Matter halos, MOOC}

\authorrunning{P. Salucci C. Antolini}

\titlerunning{ }

\maketitle


\section{Introduction}
Dark Matter is a main protagonist in our understanding of the Universe. As an 
example, let us recall the very recent observations from the 
Planck Mission in which the existence and the position of the acoustic 
peaks in the temperature spectrum leaves little option other than the existence of dark 
particles filling the entire cosmos and driving the 
formation of its structures \cite{}. It is well known that this and 
other independent observations point to a 25$\%$ of the density 
of mass energy of the Universe residing in a still unknown 
massive (particle) form.

On galactic scales, the evidence for the phenomenon we wish to 
explain in terms of Dark Matter is outstanding. The inner 
kinematics of gaactic structures so as the properties of their lensing of distant 
sources (e.g.~\citealt{rubin80,bosma81}) reveals the presence 
of something that the great majority of astrophysicists 
considers a dark ``mass component''. Furthermore, there is 
also a general acceptance that the luminous component made 
of gas and stars is playing an important role in 
determining the present day properties of galaxies including 
their dark component. This is pretty about all on the shared 
knowledge. In fact, the scientific community has no unique answers from 
universally shared truths about the 
distribution of dark and luminous matter in galaxies of 
different mass and Hubble type and how it compares with 
predictions from fundamental theories of elementary 
particles and galaxy formation. Furthermore, in investigating this 
issue even the connections themselves among phenomenology, 
simulations and theory are under debate. Let us stress that 
in the the past 30 years a quite small number of groups and 
individuals has worked on this issue. The number is 
insufficient for the knowledge transfer especially if we 
consider that this topic has become indispensable to every 
single astrophysicist and elementary particle physicist. 
Furthermore, even experts sometime cannot agree also on 
fundamental points of the topic. Today, to teach DM in 
Galaxies to 1000 classes/year and to 10000 individuals/year 
is not a very easy task.

A Massive Open Online Course (MOOC) is an online free course 
aiming at large scale interactive
participation and open access via the web. MOOCs provide 
interactive user forums that help build a
community for the students, professors, researchers. 
Although there has been access to free online courses on the 
Internet for years, recently the quality and quantity of 
courses has changed. Nowadays, the access to free courses 
has allowed students to obtain an high level of education. 
According to the New York Times ``in the past few months several ten 
thousands motivated students around the world who lack 
access to elite universities have been embracing them as a 
path towards sophisticated skills...'' \cite{} The Top 10 Sites for 
Free Education With Elite Universities include now the 
University of Stanford and Berkeley, MIT, Harvard and so on.

The question is: what about a MOOC course on Dark Matter in 
Galaxies? Of course, this must be thought as the first step 
of a fruitful collaboration between the worlds of innovative 
teaching and that of Astrophysics.
This proposal comes out by the fact that in our opinion, in 
addition to the advantages discussed above, a MOOC on Dark 
Matter in Galaxies will allow 1) a more general uniformity 
and better quality of the content taught to students 
everywhere 2) the formation of a world community that includes scientists 
active in this field and, e.g., students in countries where 
Astrophysics is not developed. 3) an impulse for the 
scientific community to search for a shared knowledge.

What does the reader think about it? Discussion about MOOCs in many 
countries has not yet started, personally a month ago we 
never heard of them.
We started to grasp the potential of MOOCs for Cosmology as 
we heard of the MOOC Production Fellowshipâs contest, hosted 
by Stifterverband fur die Deutsche Wissenschaft and 
iversity. The ten winning projects will receive funds and 
technical help to implement their MOOC concepts. In our 
particular case, to win one of the ten assigned fellowship is 
to speed up innovative teaching of astrophysics into (a 
possible) future because we believe that, today, in this 
field, without appropriate financial support, the will   
 to produce such courses is  near to zero.

The SISSA project ``Dark Matter in Galaxies'' envisages 25 
one-hour lectures of highly fruitfully interaction between students and lecturer. 
Original data and codes will be made available for the learning 
pipeline. A global forum will be set up for ideas exchange, discussion, feedback and insights from experts in the field. This project has been presented by one of us at Euclid Consortium  Leiden Meeting in the "Education and Public Outreach"  splinter session,  receiving a good interest 

Readers that want to see details of this project and even  vote for it,
 please follow the link

https://moocfellowship.org/submissions/dark-matter-in-galaxies-the-last-mystery

We thank of course, those  the vote for it (the interest of the general public is very 
important in order to win the fellowship and this project is the only one presented in the wide field of astrophysics).

Of course small or even negative interest for what was 
discussed here, means no vote! Such a feedback would be useful to 
set the present point of view of the scientific community.

\begin{figure}

\centerline{ }

\caption{ }

\end{figure}

\end{document}